\RequirePackage{snapshot}
\documentclass{elsarticle}

\usepackage{lineno,hyperref}
\modulolinenumbers[5]

\usepackage{listings}
\usepackage[utf8]{inputenc}
\usepackage{graphicx}
\usepackage{subcaption}
\usepackage{xcolor}

\biboptions{authoryear}

\begin{document}

\begin{frontmatter}

\title{An interactive sequential-decision benchmark from geosteering}


\author[norce]{Sergey Alyaev\corref{mycorrespondingauthor}}
\cortext[mycorrespondingauthor]{\textsuperscript{\textcopyright} 2020. This manuscript version is copyright to the authors and is made available under the license \url{http://arxiv.org/licenses/nonexclusive-distrib/1.0/}. Corresponding author}
\ead{Sergey.Alyaev@norceresearch.no}

\author[uis]{Reidar Brumer Bratvold}

\author[bend]{Sofija Ivanova}

\author[norce]{Andrew Holsaeter}

\author[bend]{Morten Bendiksen}


\address[norce]{NORCE Norwegian Research Centre, Postboks 22 Nygårdstangen, 5838 Bergen, Norway}
\address[uis]{University of Stavanger, Postboks 8600 Forus, 4036 Stavanger, Norway}
\address[bend]{Bendiksen Invest og Konsult, Fredlundveien 83B, 5073 Bergen, Norway}

\begin{abstract}

During drilling to 
maximize future expected production of  hydrocarbon resources the experts commonly adjust the trajectory (geosteer) in response to new insights obtained through real-time measurements.
Geosteering workflows are increasingly based on the quantification of subsurface uncertainties during real-time operations. 
As a consequence operational decision making is becoming both better informed and more complex.
This paper presents an experimental web-based decision support system, which can be used to both 
aid expert decisions under uncertainty or 
further develop decision optimization algorithms in controlled environment.
A user of the system (either human or AI) controls the decisions to steer the well or stop drilling. 
Whenever a user drills ahead, the system produces simulated measurements along the selected well trajectory which are used to update the uncertainty represented by model realizations using the ensemble Kalman filter. 
To enable informed decisions the system is equipped with
functionality to evaluate the value of the selected trajectory under uncertainty with respect to the objectives of the current experiment.

To illustrate the utility of the system as a benchmark,
we present the initial experiment, in which we compare the decision skills of geoscientists with those of a recently published automatic decision support  algorithm.
The experiment and the survey after it showed that most participants were able to use the interface and complete the three test rounds.
At the same time, the automated algorithm outperformed 28 out of 29 qualified human participants. 

Such an experiment is not sufficient to draw conclusions about practical geosteering, but is nevertheless useful for geoscience.
First, this communication-by-doing made 76\% of respondents more curious about and/or confident in the presented technologies.
Second, the system can be further used as a benchmark for sequential decisions under uncertainty. This can accelerate development of algorithms and improve the training for decision making.


\end{abstract}

\begin{keyword}
interactive benchmark
\sep
sequential geosteering decisions
\sep
uncertainty quantification
\sep
expert decisions
\sep
experimental study
\sep
decision support system
\end{keyword}

\end{frontmatter}


\newcommand{\todo}[1]{}

\newcommand{\TODO}[1]{\todo{#1}}

\newcommand{\dss}{DSS-1}
\newcommand{\dui}{GUI}
\newcommand{\bestInc}{HP-01}
\newcommand{\bestCon}{HP-06}
\newcommand{\bestRel}{HP-08}

\todo{change to a consistent name}


\section{Introduction}
Geosteering is the intentional control of a well trajectory based on the results of down-hole real-time geophysical measurements \citep{Shen2018}.
Traditionally, research in geosteering has been focused on interpretation of log measurements. 
As a result during the last decade there has been a steady growth of automated methods for the measurement inversion and interpretation which yield steadily growing amounts of data that needs to be handled by the decision makers.
This data opens the possibility to target the oil-bearing zones which were not economically viable previously \citep{Larsen2016}. 
At the same time this also makes the decision-making more complex by adding more relevant information to consider and evaluate in real time
\citep{Hermanrud2019}.

The literature review in 
\cite{Kullawan2016value}
showed that there was hardly any prior publication that considered a consistent framework for geosteering decision making with several objectives. 
The authors prepared an alternative decision-focused approach. In the last few years, there have been several more attempts to address geosteering as a sequential decision problem. 
\cite{Kullawan2014}  introduced a multi-criteria framework optimized for sequential decisions in geosteering. 
\cite{veettil2020bayesian} developed a Bayesian estimator of stratigraphy that can be further extended with forward well planing. 
\cite{Chen2015spe,Luo2015} considered an ensemble-based method for optimization of reactive steering under uncertainty. 
\cite{Kullawan2018} demonstrated application of dynamic programming for finding optimal long-term decision strategies for a certain set of geosteering problems. 
In \cite{alyaev2019dss}, a simplified dynamic programming algorithm was used in a context of a more general geosteering problem with several targets.
\cite{kristoffersen2020unpublished} proposed an reinforcement-learning-based approach to steering based on the initial field planning. 

Specific to optimization of sequential decisions under uncertainty is that the acquired data and hence the uncertainty for subsequent decisions will depend on the previous decisions. 
For geosteering, the data is acquired along the well path and will hence depend on the selected drilled trajectory. 
Therefore the papers reviewed above relied either on a static benchmark   sacrificing uncertainty updates 
(\cite{kristoffersen2020unpublished} used Olympus) 
or needed to develop a testing platform alongside the decision methodology \citep{Chen2015spe,Luo2015,Kullawan2018,alyaev2019dss}.
On one hand, the descriptive optimization benchmarks either 
contain static pre-defined environment which is not influenced by
decisions (e.g. Olympus, see \cite{fonseca2018overview}) 
or 
are ambiguous due to open choice of both updating and decision strategies
(e.g. Brugge, see \cite{peters2013extended}).
On the other, the benchmark tool-kits for developing and comparing sequential decision agents, such as OpenAI Gym \citep{brockman2016openaigym} are primarily targeting reinforcement learning  for games and simple physical simulators.
Thus, there is currently a lack of benchmark systems providing an uncertain geological environment where the uncertainty changes consistently in response to taken decisions.

For this study we have developed a web-based platform which can 
update a 2D geological model in response to decisions and share the current state of the system via an API and a GUI.
Thus the platform serves not only as a benchmark for decision-agent development, but also a tool which can compare algorithms to decisions of experts.
We are using the problem set-up of multi-target geosteering from \cite{alyaev2019dss} which is simple to explain, yet hard to master for either humans or robots. 
We represent uncertainty by an ensemble of realizations which are updated consistently by the Ensemble Kalman Filter (EnKF) based on synthetic electromagnetic measurements  \citep{Chen2015spe} along the selected well path. 
For the GUI we also developed new visualization shows how the geometric uncertainty relates to expected value of the planned well. 

We present the first experiment of use of the web-based platform, 
in which 
formation evaluation and geosteering experts competed with 
fully automated system from \cite{alyaev2019dss}
for getting the highest well value.
The purpose of the experiment was to compare the decisions of the experts with the fully automated algorithm.
At the same time it facilitated learning-by-doing
communication of concepts related to handling of uncertainty decision theory and optimization related to geosteering.
We feedback from the experiment also gave input to further development of the platform.



The paper is organized as follows: First, we describe the experimental set-up which is currently used for the platform.
After that, Section \ref{sec:platform} presents the GUI and the API available to the users of the platform (the corresponding code is linked in Section \ref{sec:repo}).
In Section \ref{sec:results} we present the first experiment run on the platform, including the results of the expert and the feedback that we have received.
Finally, the findings of the paper and further perspectives for the platform usage are summarized in Section \ref{sec:conclusions}.

\section{Description of the experiment}
\label{sec:rules}
In this section we describe the set-up of the geosteering decision-making experiment that runs in our sandbox environment. 

\subsection{Objective}
\begin{figure}
    \centering
    \includegraphics[width=0.5\textwidth]{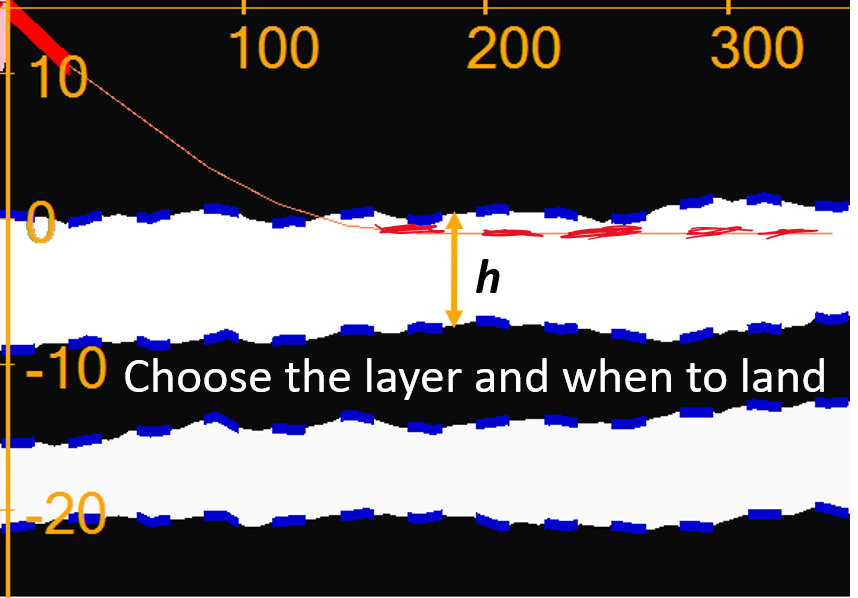}
    \caption{An example of synthetic truth and a possible steering trajectory. The highlighted part of trajectory gives positive score.}
    \label{fig:two_layers_h}
\end{figure}

An experiment is split into rounds.
The objective for the decision-making task in each round is to make landing and steering decisions in a multi-layer geological setting. 
The pre-drill model contains 5 alternating layers: shale-sand-shale-sand-shale. 
The goal is to maximize an approximate NPV of the well. 
This is done by landing and staying near the roof of a sand layer with considerations of layer thickness and drilling costs. More specifically, the objective score is calculated by the following rules: the participant gets:
\begin{itemize}
    \item $h$ points for every meter in sand layer (along X-axis), where $h$ is the layer thickness
    \item $2*h$ points when they drill in the sweet-spot near the roof (0.5 m to 1.5 m from the top boundary of a sand)
    \item negative $c$ points is the cost of drilling every meter, where $c = 0.086$.
\end{itemize}
An example of synthetic truth and a possible steering trajectory is shown in Figure \ref{fig:two_layers_h}.

The decision locations are evenly spaced along the X-axis.
Each round of a competition consists of at most 14 geosteering decisions, each being either a change in the direction of the next drill-stand or a stopping decision

\subsection{Uncertainty and updates}
\label{sec:update}
Like any decision, geosteering decisions are made under uncertainty. 
The main uncertainty during geosteering is the lack of complete knowledge of the geology through which the well will be drilled. 
We represent uncertainty using an ensemble of 120 realizations of the layered geology in 2D. 

The original ensemble is based on a prior distribution which is also used to generate the synthetic truth in the experiment. The layer boundaries are discretized along X-direction and Y-position for every boundary is generated using a variogram model with kriging as described in \cite{alyaev2019dss}.

The ensemble of realizations is updated following each decision using the Ensemble Kalman Filter (EnKF) algorithm described in \cite{Luo2015}. 
\footnote{The EnKF is a Monte-Carlo (discrete) approximation of the Kalman Filter. It gives an approximation of a Bayesian update with Gaussian priors and likelihoods.}
For the update we use measurements produced from a synthetic EM tool which is located at the drill-bit and has look-around capability of  4.8 meters up, down, and sideways. 
The model for the tool is described in \cite{Chen2015spe}.
The system performs one update between decision points which uses measurements in three equally distributed locations.

\section{Description of the web-based platform}
\label{sec:platform}
One of the main purposes of the developed web-based platform is to enable comparison between experts and algorithms when it comes to decision-making for geosteering. The developed web-based platform includes both a GUI for experts and an API for AI bots. 
In this section we first describe the capabilities of the platform through its GUI and then explain how corresponding API can be used by a bot. 
We finish the section by a summary of the implementation of \dss\ from \cite{alyaev2019dss}.

\subsection{The GUI}
\label{sec:gui}
\begin{figure}
    \centering
    \includegraphics[width=0.9\textwidth]{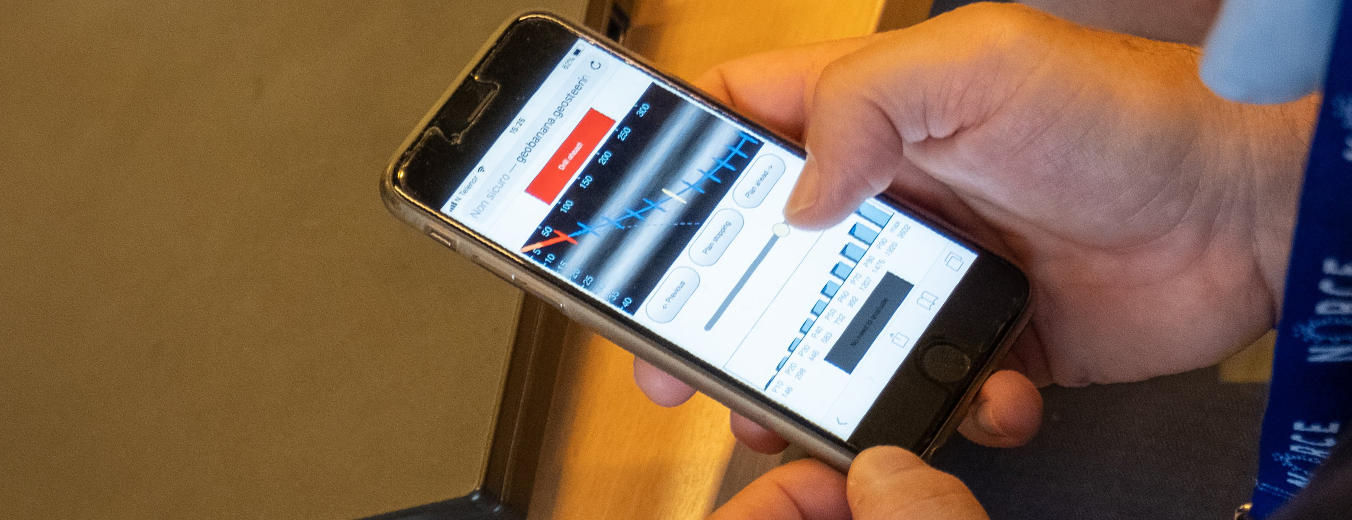}
    \caption{The \dui{} running on a mobile phone during the experiment.}
    \label{fig:fullInterfaceMobile}
\end{figure}

To evaluate the decision-making strategies of the experts, we developed a simple online decision-support Graphical User Interface (\dui). 
The online mobile application  
gives the contestants the same information that a geosteering decision algorithm would get wrapped into a user-friendly \dui, see Figure  \ref{fig:fullInterfaceMobile}.

\subsubsection{Uncertainty visualization}
\begin{figure}
    \centering
    \includegraphics[
    trim={0 5.3cm 0 0},clip,
    width=0.95\textwidth]{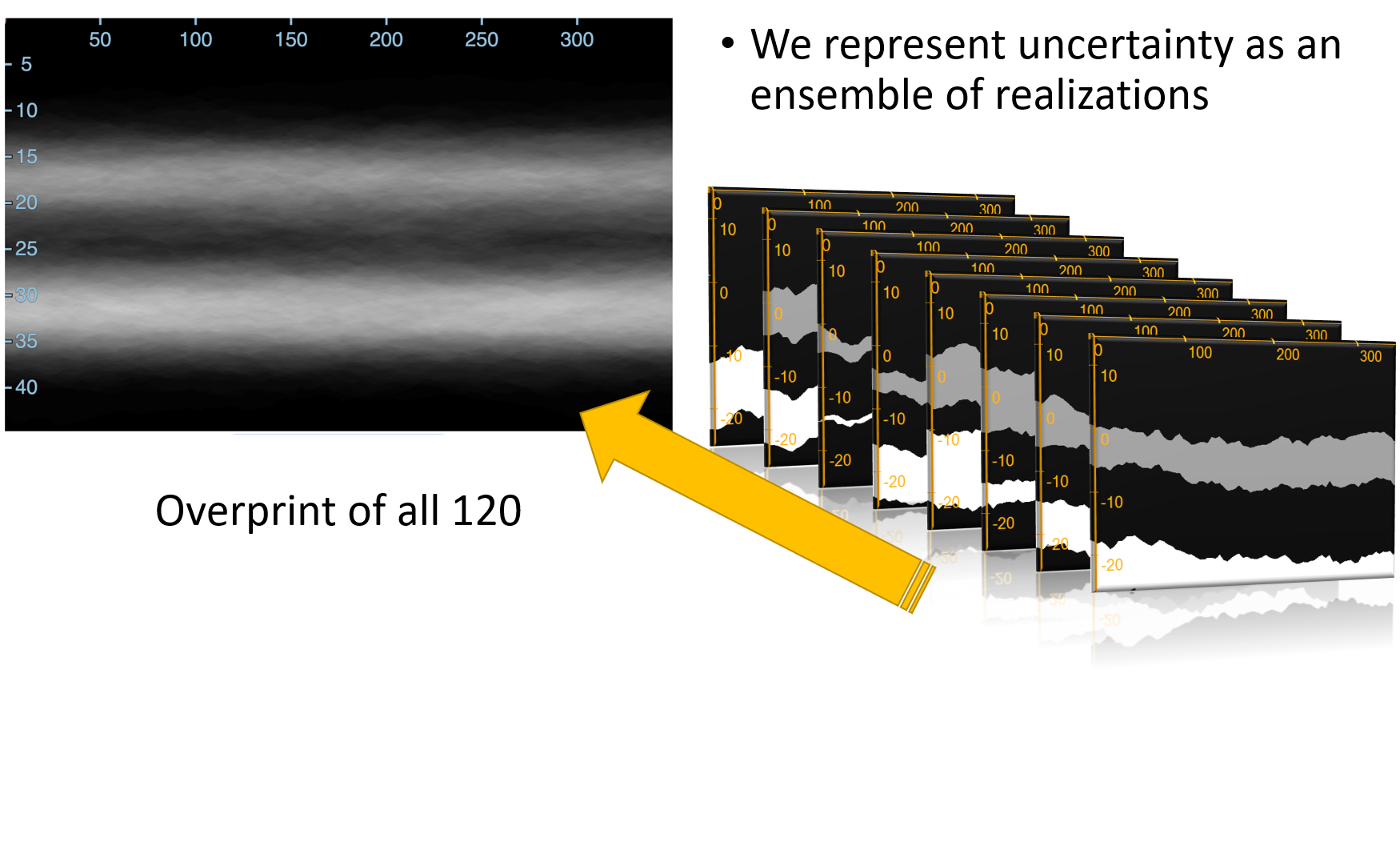}
    \caption{Uncertainty is represented by an ensemble of realizations and visualized as an overprint of all ensemble members.}
    \label{fig:overprint}
\end{figure}

One of the primary sources of information for the system users is the graphical display of the uncertain earth model. 
The users can view an overprint of the ensemble which provides a display of the (white)  sand layers' location uncertainties (Figure \ref{fig:overprint}).

\subsubsection{Planning and committing to a decision}
\begin{figure}
    \centering
    \includegraphics[width=0.5\textwidth]{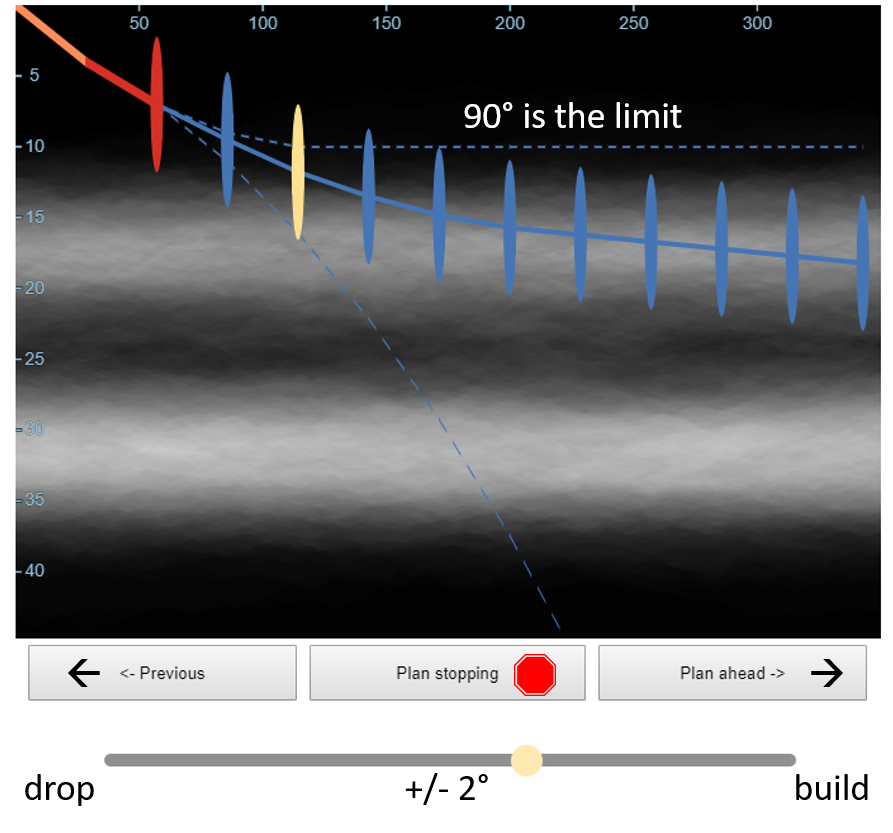}
    \caption{A user interface showing controls for steering and steering limits. The ellipses represent decision points and their size indicates the look around of the EM tool. The orange part of trajectory is already drilled; the red part is the next decision to commit to; the blue part is the plan ahead. The yellow ellipse shows the selected point at which the steerer can adjust the dip. The selection can be moved by the buttons.}
    \label{fig:interface}
\end{figure}

The discrete locations of the geosteering decisions to be made in this round are shown graphically in Figure \ref{fig:interface}.


Until the well is finalized, the steerer can plan ahead the entire well by changing the dip of the well in the decision points 
(ellipses in Figure \ref{fig:interface}).
Alternatively, the contestant can decide to stop drilling at any of the points. 
The latter might be optimal if, for example, the well entered the underburden.

At each decision point the participant must commit to a decision: choose whether to adjust the dip for the next drill-stand or stop drilling.
Stopping decisions implies that the well is finalized, and no further drilling steps can be taken. 

\subsubsection{Tools to make informed decisions}
To aid in their decision making, the contestants are presented with a visual decision support tools in the \dui. 
The \dui{} dynamically updates uncertainty as the well is drilled and helps to estimate the well value. 

\begin{figure}
    \centering
    \includegraphics[
    trim={0 1cm 0 12cm},clip,
    width=0.6\textwidth]{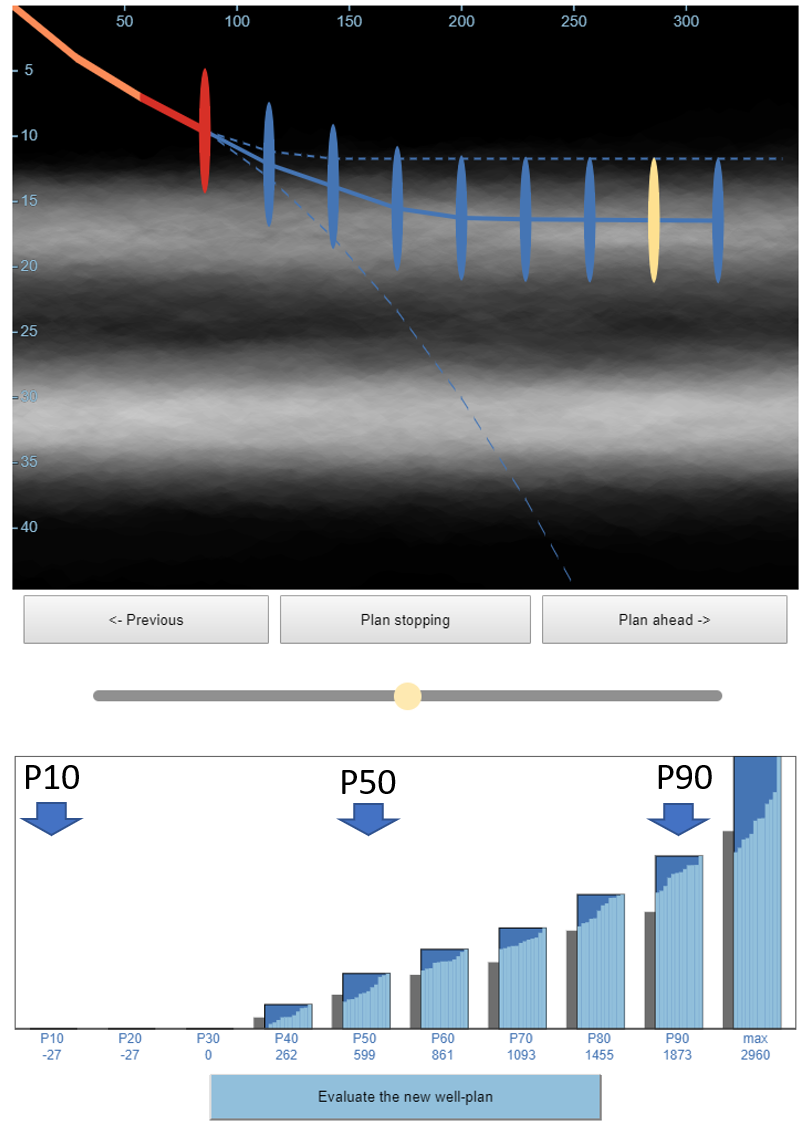}
    \caption{A score distribution diagram shown to a user based on current set of realizations representing the uncertainty.}
    \label{fig:distributionVisualization}
\end{figure}

Once a well trajectory is planned, it can be evaluated using the scoring function with respect to the current understanding of uncertainty (the ensemble). 
The results of this evaluation are summarized in a bar diagram as shown in Figure \ref{fig:distributionVisualization}. 
The diagram shows a cumulative density diagram based on the 120 ensemble members (light blue). 
The results are grouped into percentiles of value (P10 - P90) shown in dark blue. 
The interface also shows the percentile values for the previous evaluation as gray bars on the background (Figure  \ref{fig:distributionVisualization}).

\begin{figure}
    \centering
    \includegraphics[width=0.5\textwidth]{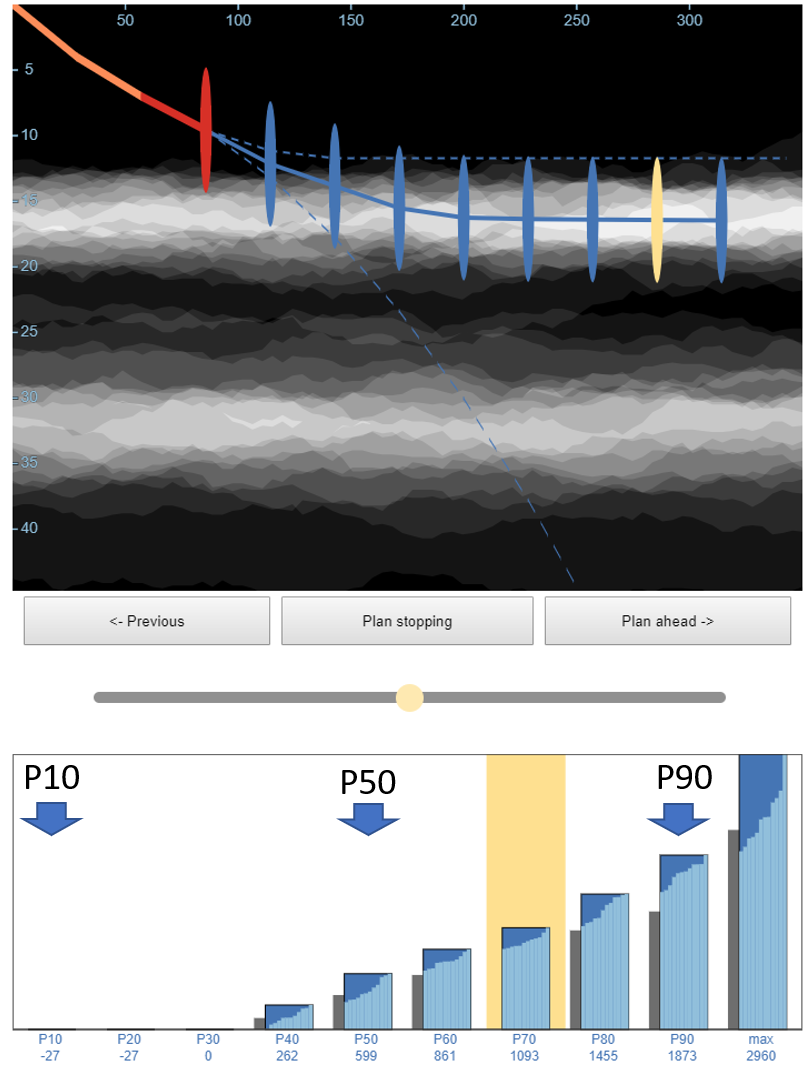}
    \caption{An example of selecting subset of realizations using the interactive score distribution diagram. }
    \label{fig:subsetOfRealizations}
\end{figure}

The percentile / cumulative density diagram is interactive. 
The user can select a percentile to see the subset of realizations that give the selected value range, e.g. between P60 and P70  
(Figure  \ref{fig:subsetOfRealizations}).

\subsection{The API}
The GUI described above exposes the functionality of a the workflow 
from \cite{alyaev2019dss} to human users.
In this section we describe the API of the system which is exposed via the web platform.

\subsubsection{Uncertainty communication}
The uncertainty is communicated to the system using the native format of the ensemble-based earth model. The system transmits an array of 120 relizations, each containing the grid of X-positions in an ascending order together with four boundaries represented as a sequence of corresponding Y-positions. 
The first two boundaries is the roof and the floor of the top sand layer, and the second two correspond to the bottom layer.

\subsubsection{Committing a decision}
For each sequential decision point the API gives the trajectory that has been already drilled.
A bot needs to commit to a decision
of either continuing or stopping. 

The continuing decision should be represented by the coordinates of the committed point $(x,y)$, where $x$ is fixed for a decision step and $y$ should be fulfil the limit on the dog-leg-severity angle (+/- 2 degrees).
After the decision to continue, the API will return the updated realizations based on the acquired data along the drilled segment. 

After stopping, the system receives the score obtained within the synthetic truth as well as the rating relative to other participants of the round and has possibility to request a new round. The actual score can be used for training of data-based algorithms.

\subsubsection{Making informed decisions}
Similar to human participants, a bot can request evaluation of a score for a given trajectory.
Given a sequence of $(x,y)$ pairs, the API returns evaluation of the given trajectory represented by a sequence of points for each member in the ensemble of realizations. The response is an array of pairs, each consisting of the score and the index of realization for which that score was observed.

\subsection{Implementation of \dss}
\label{sec:dss}
The web-based platform was developed from the functionalities that have been used by \dss\ described in \cite{alyaev2019dss}.
The update part of the \dss workflow\ is used by the web platform directly and was described in Section \ref{sec:update}. 
Here we summarise the optimization algorithm of \dss. 

At every decision point the optimization is divided into two steps: global deterministic optimization for each realization and robust optimization for the immediate decision.

On the first step \dss\ considers all discrete trajectories until the end of the operation. 
The discretization is made so that for the decision step $x_0$ the system uses a finite number of $y_i$ distributed on a selected regular grid. 
For every of these trajectories the system stores the score of the well for each of realizations. 
In this paper we use a version with a discount factor for future decisions. 
That is, every next segment's value is multiplied by 0.9 and thus has influence on the score of the well, which compensates for uncertainty of the future learning.
The details about the efficient implementation of \dss are presented in \cite{alyaev2019dss}.

On the second step \dss\ considers all discrete alternatives for the next (immediate) decision, i.e. all $(x_1,y_i)$ within the geometric constraints and stopping and chooses $y_{i}^{\textrm{opt next}}$, which gives best decision on average:

\begin{equation}
   y_{i}^{\textrm{opt next}} = \arg\max_{y_{i}} 
   \frac{1}{n}\sum_{j=1}^{n} 
   \left\{
   O([(x_0,y_0), (x_1,y_i)]|M_j) 
   \right.
   +\left.
    \gamma O([(x_1,y_i),...]|M_j)
   \right\},
   \label{eqOptimizationGlobal}
\end{equation}
where $n=120$ is the number of realizations, $O(\psi|M_j)$ is the objective function computed for trajectory $\psi$ against realization $M_j$, $[(x_0,y_0), (x_1,y_i)]$ is a segment from current point to the next decision point ending in $y_i$, $O([(x_1,y_i),...]|M_j)$ is the optimal value for trajectory starting in $(x_1,y_i)$ in the realization $M_j$ computed in the first step, and $\gamma$ is the discount factor. 
In this we use $\gamma = 0.9$.

Based on the description above we can summarize the following properties of the \dss\ decisions:
\begin{itemize}
    \item The \dss\ optimization algorithm described above gives is deterministic function of the discretization, realizations, and the objective function. 
    \item The \dss\ only uses the evaluation of the objective function and not the representation of the earth model. This makes it flexible with respect to earth model implementation and the objective function. At the same time it does not use all the information available in the experiment.
    \item The global optimization used by \dss\ requires up to 100000 evaluations of objective function for every decision which is impossible for a human, but can be performed within seconds on a computer.
\end{itemize}

\section{Results and discussion}
\label{sec:results}
The well-based platform has been developed to compare the \dss{} to human experts and, through this comparison, to communicate the concepts related to real-time decision making under uncertainty. 
In this section we first present the result of the first experiment 
which was held as a plenary session of the Formation Evaluation and Geosteering Workshop 2019 by NFES and NORCE held in Stavanger Norway \citep{nfes2019}.
We evaluate decision quality of experts and \dss\ based on the collected data.
After that we show the results of the survey among the participants of the experiment which addresses the usefulness of such an experiment in communication of the research.
Finally we summarise feedback about the individual components of the web-based platform and its usefulness for practical training. The latter should direct further research on system-user interaction.

\subsection{The first experiment}
\label{sec:experiment}
The rounds of the experiment were organized as follows.
After the presentation of the rules (Section \ref{sec:rules}) and the interface (Section \ref{sec:gui}), the contestants had a practice period of 15 minutes to familiarize themselves with the competition set-up. 
During this period a \dui{} expert was showing the usage of the features of the user interface and a possible strategy for geosteering on a big screen.

\begin{figure}
    \centering
    Round 1\\
    \includegraphics[width=0.8\textwidth]{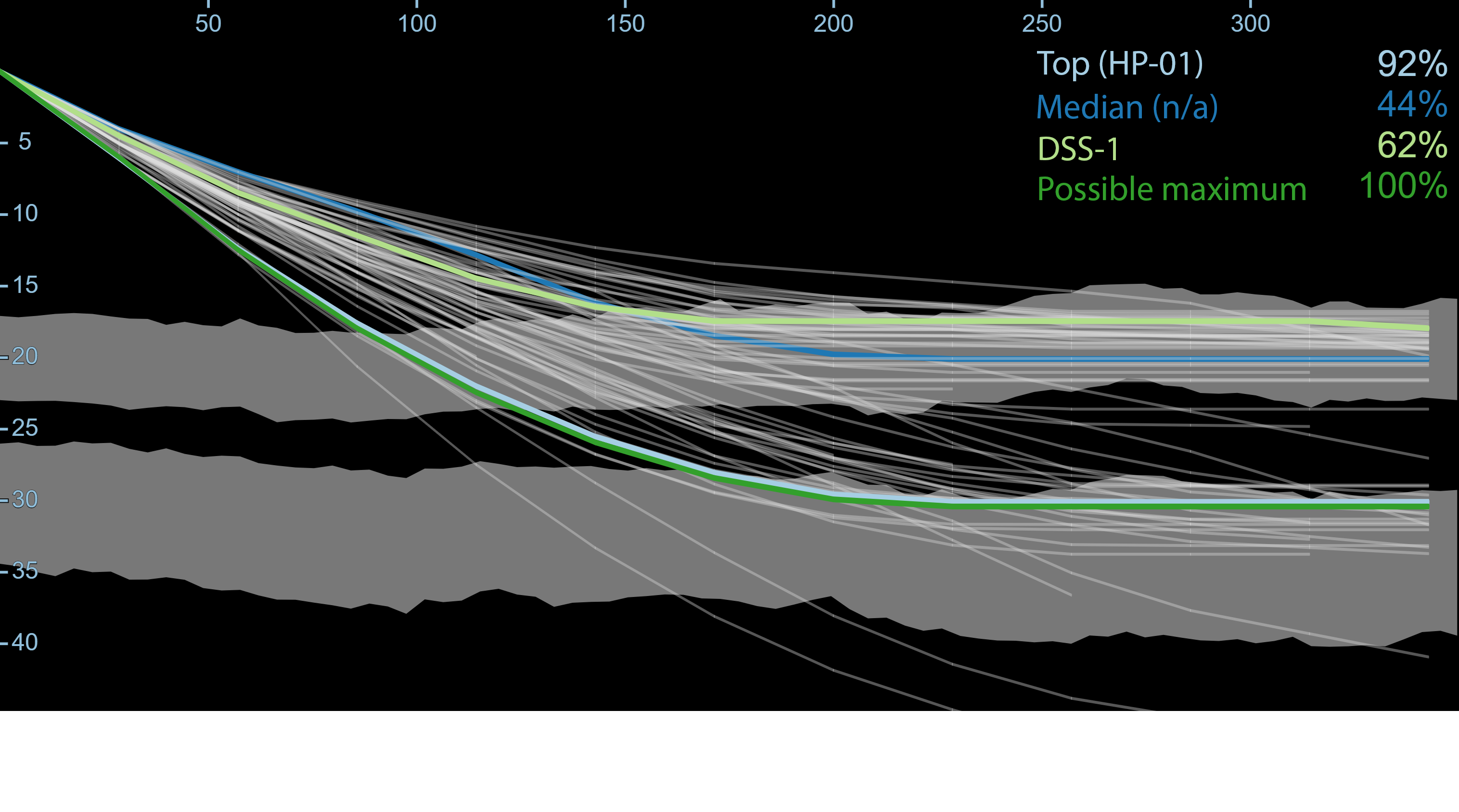}
    \\Round 2\\
    \includegraphics[width=0.8\textwidth]{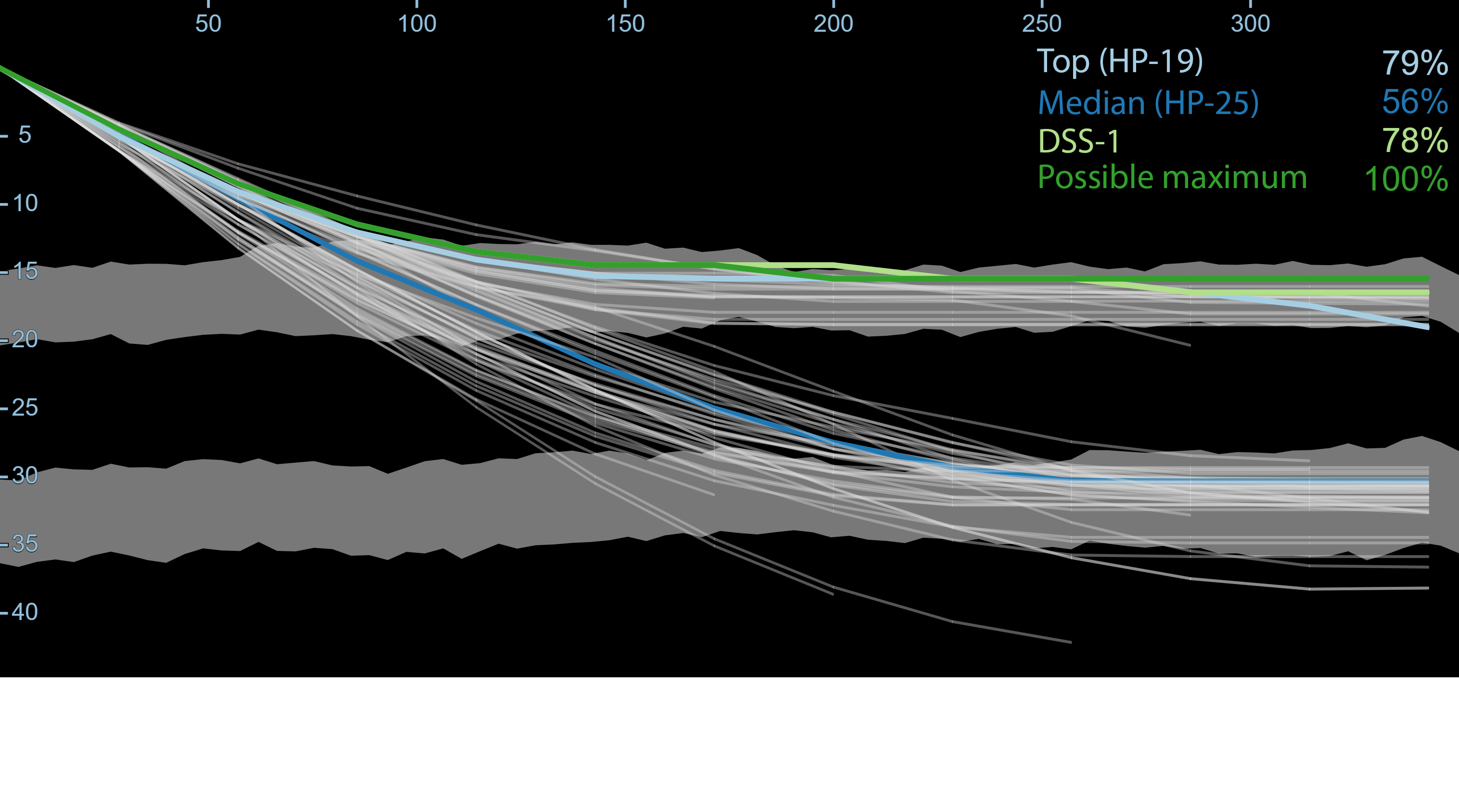}
    \\Round 3\\
    \includegraphics[width=0.8\textwidth]{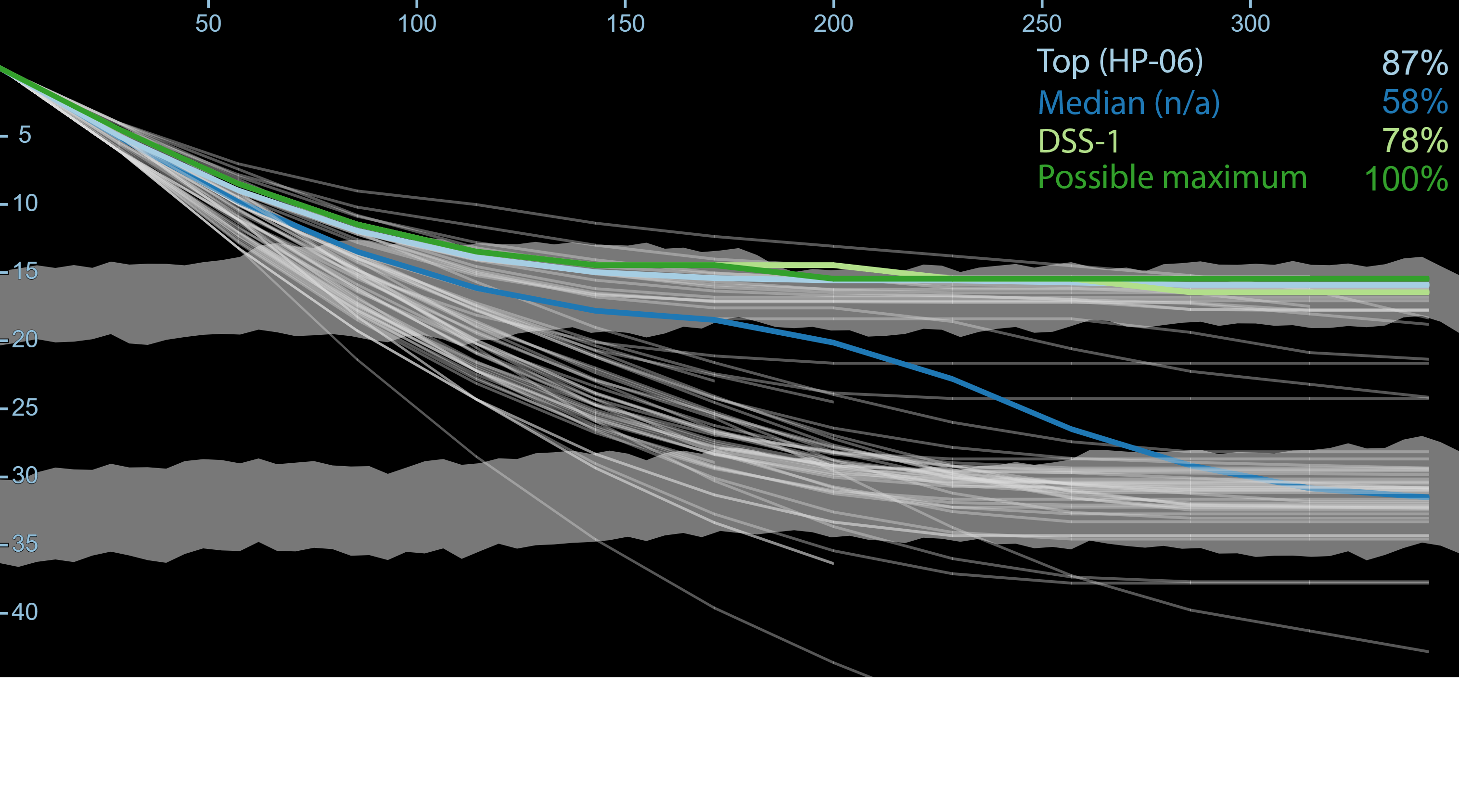}
    \caption{The well trajectories drilled by participants in each of the three rounds (also including not qualified participants). The highlighted trajectories show: the top participant result; the median participant result; \dss\ result; and the solution obtained by optimization assuming perfect information (Possible maximum). For the qualified participants the participant ID is shown in the paranthesis.  
    }
    \label{fig:results}
\end{figure}

Following the demonstration, there were three scoring rounds of approximately 6 minutes each. 
All rounds had an identical ensemble of starting realizations but  a different synthetic truth unknown to the participants.  
To evaluate consistency of decisions, the truths were chosen as: 
\begin{itemize}
    \item Round 1) The bottom layer was optimal
    \item Round 2) The top layer was optimal
    \item Round 3) Identical to Round 2, to allow comparison of the consistency of contestants' decisions under the same conditions.
\end{itemize}
The synthetic truths as well as the optimal solution computed by deterministic optimization on the synthetic truth are shown in Figure \ref{fig:results}. 

Out of the 75 workshop participants, 55 participated in all three rounds. 
The wells 'drilled' by all the participants are compared with the optimal trajectory in Figure \ref{fig:results}.
A fraction of the 55 participants did not reach any of sand layers or were affected by software issues. 
For fairness, we disregard them from the results, and  consider the remaining 30 participants, whom we call \textbf{qualified participants}.
Among the qualified participants was \dss\ described in Section \ref{sec:dss}.

\subsubsection{Analysis of the results}








Decision analysis
defines a good decision as the one that is
logically consistent with the alternatives (steering choices), information (representation of geological uncertainty), and values (objectives) brought to the decision \citep{bratvold2010making,abbas2015foundations}.
In the decision analysis process, the outcome of a single decision does not imply the quality of that decision.
That is, given the uncertainty, a good decision may lead to a bad outcome and vice versa. 
Decision analysis framework allows to identify good decisions before knowing their outcome by recording the principal inputs of the decision-making process and the corresponding decision strategies.


The decision strategy of \dss\ (see Section \ref{sec:dss}) is designed on the principles of the robust optimization \citep{Chen2015spe,alyaev2019dss}, which are known to lead to better decisions under uncertainty.
The decision strategies of human participants, however, are not so easy to deduce and to analyse.
One possible approach is to survey participants about their strategies \citep{welsh2005cognitive,alyaev2021systematic}.
One can argue, that such approach is subjective and reported strategies do not always translate into practical decisions.
The ambition of the proposed experiment is to assess the quality of decision strategies of the participants based on the decision-outcome data recorded in a send-box environment.


Clearly, the geosteering experts could outperform an automated system if they possess knowledge and/or logic over and beyond what is built into an automated system. 
Thus we expect that the experts would outperform the system in a real-world geosteering context.
However, in this controlled experiment, we put the decision makers and the system in equal conditions in terms of information availability. 
We admit, however, that the results might not objectively reflect the  decision strategies for real geosteering because the controlled environment was not familiar to the experts and since the competition format might influence the participants' behaviour.

The outcome of every decision is a result of both skill and chance. Therefore, it is quite possible to achieve good results over the 14 decisions made for one well with a weak strategy. 
However, in the long run, the decision outcomes should be representative of the decision quality of each participant. 
To reduce the influence of chance, the experiment  included a training round and three qualifying rounds.

\subsubsection{Ranking}

There is no unique method to assess results of a competition over several distinct rounds as it requires scaling the results by a chosen metric.
As a primary simple metric, we used the percentage of maximal possible result for each round which was averaged to give the final ranking.
The results are scaled by the 100\% result, which is obtained by discrete optimization on the synthetic truth for each round, thus representing a close approximation to theoretical possible maximum.
The data of the expert experiment  from the experiment including scoring is available in the linked repository, see Section \ref{sec:repo}.
The human participants are identified as HP-$n$, where $n$ is the rank (1 to 30) according to this metric.
The fully automated decision system performed better than 93\% of the participants placing 2nd among the 30 qualified participants. 
This high ranking should not be surprising given the earlier discussion in this section.

\begin{figure}
    \centering
    \includegraphics[width=0.7\textwidth]{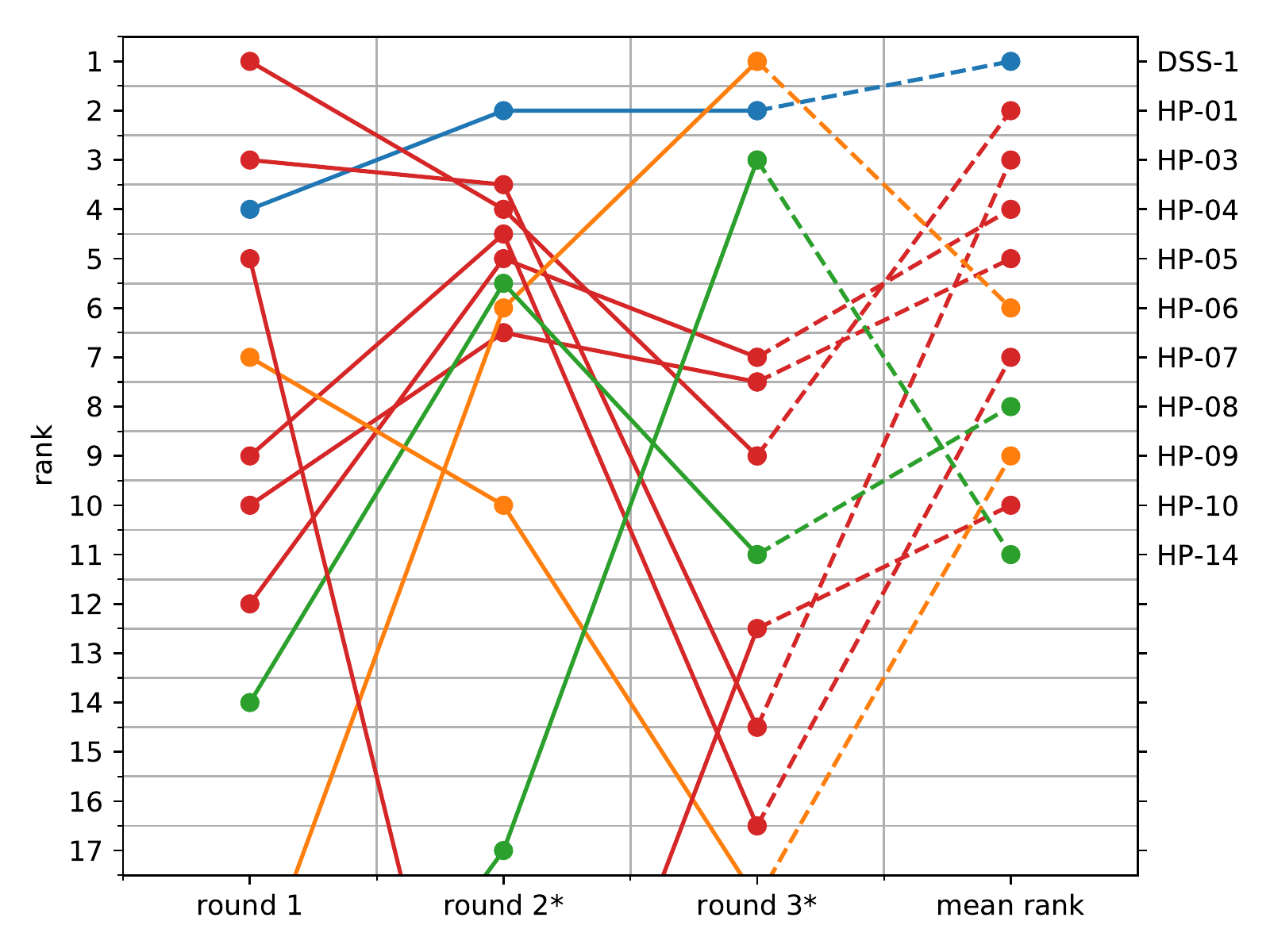}
    \caption{The comparative ranking of the top participants over three rounds of the experiment. 
    The rounds 2 and 3 followed the identical setup which enabled to derive rank which included participants' results from both rounds (60 results). 
    The rank* is derived by scaling this rank to 30 participants, resulting in fractional values.
    The mean rank is the rank (1 to 30) based on the mean of the three rounds.
    For convenience this figure uses the same color-coding as Figure \ref{fig:consistency}.
    }
    \label{fig:rankLevels}
\end{figure}

Another possibility to compare the results of different rounds is to consider the ranking within the population. 
The rank in the population is the position of the participant in the round among the other participants relative to the total size of the population.
In the case of our experiment, we take advantage of two identical rounds and arrive at rank*, common for rounds 2 and 3.
This type of ranking of top 11 participants is shown in Figure \ref{fig:rankLevels}.

While \dss\ was second in the simple ranking discussed previously 
it gets the top position in this alternative ranking. 
The simple ranking is highly influenced by results in a single round.
HP-01 got a near-perfect score (92\%) in round 1, making him the top simple-ranked participant.
At the same time neither HP-01, nor any other participant has beaten \dss\ in more then one round, bringing \dss\ to the top of comparative ranking.

As we see, comparative ranking is more objective as it reduces the influence of chance from a single round. Therefore we are planning to adopt comparative ranking as primary for future experiments.




\subsubsection{Consistency of decision strategies}

\begin{figure}
    \centering
    \includegraphics[width=0.8\textwidth]{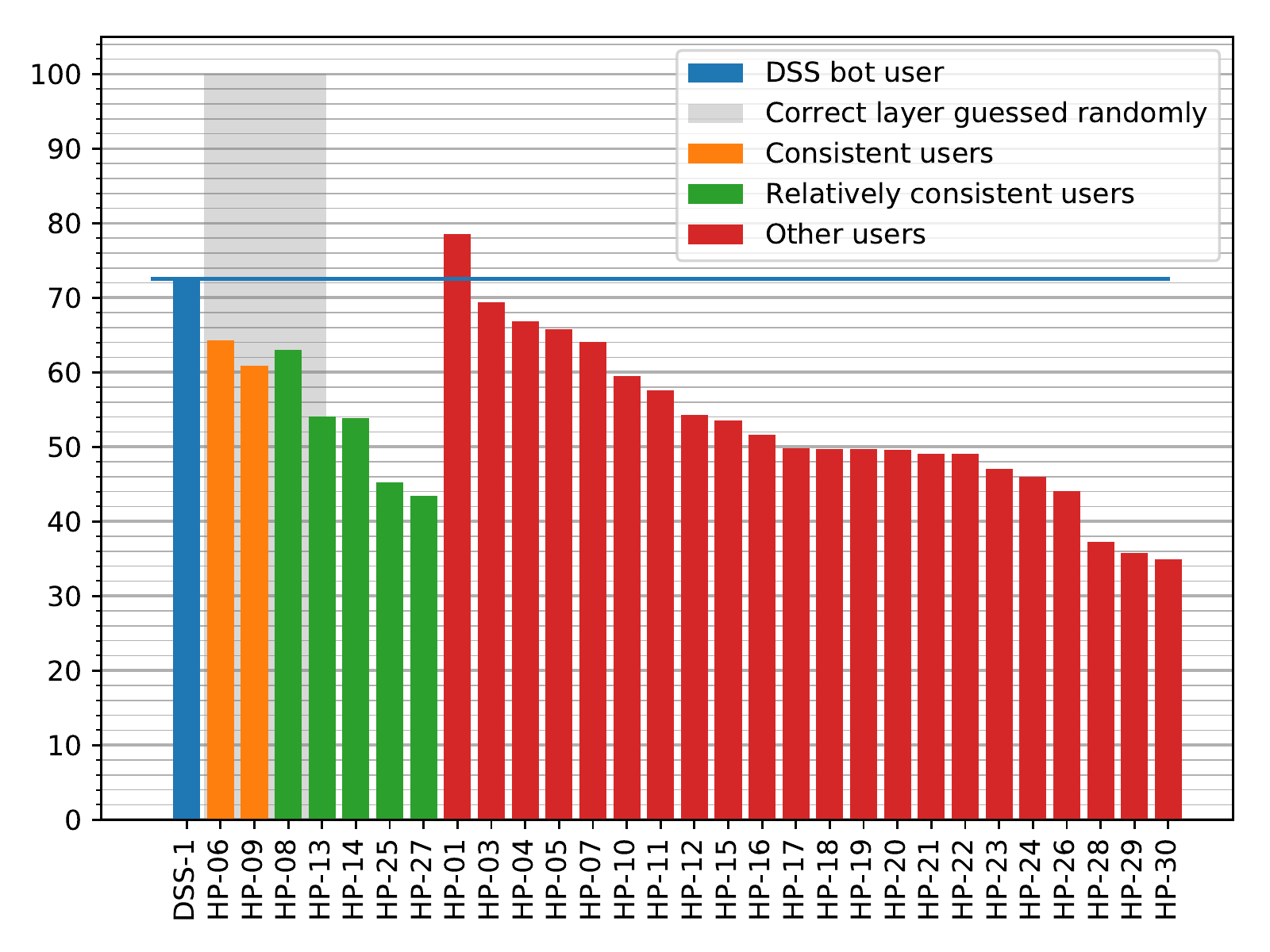}
    \caption{Consistency of decisions when performing the same task for all the participants: DSS and Human Participants (HPs).}
    \label{fig:consistency}
\end{figure}

\begin{figure}
    \centering
    a.
    \includegraphics[width=0.4\textwidth]{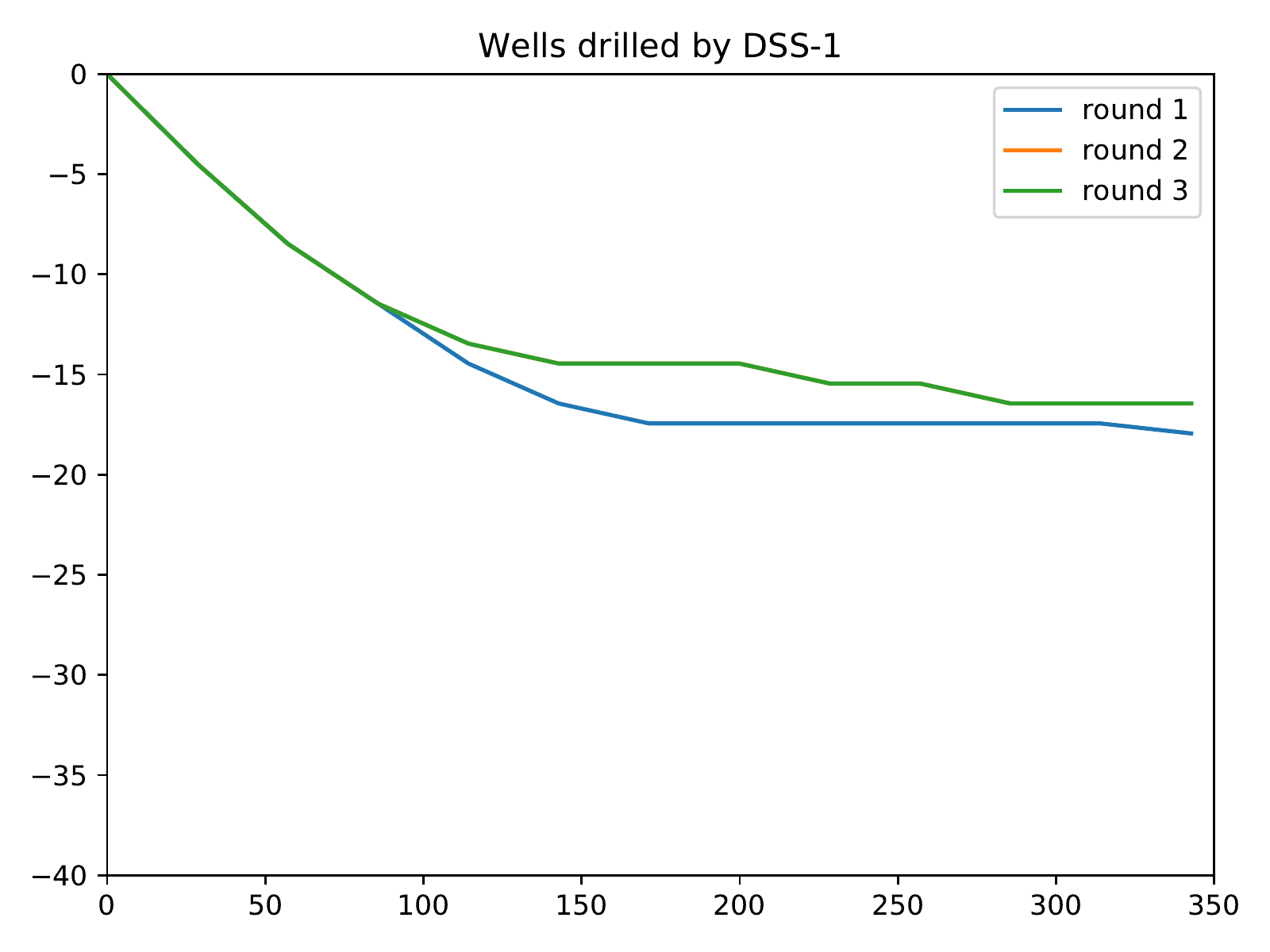}\,
    b.
    \includegraphics[width=0.4\textwidth]{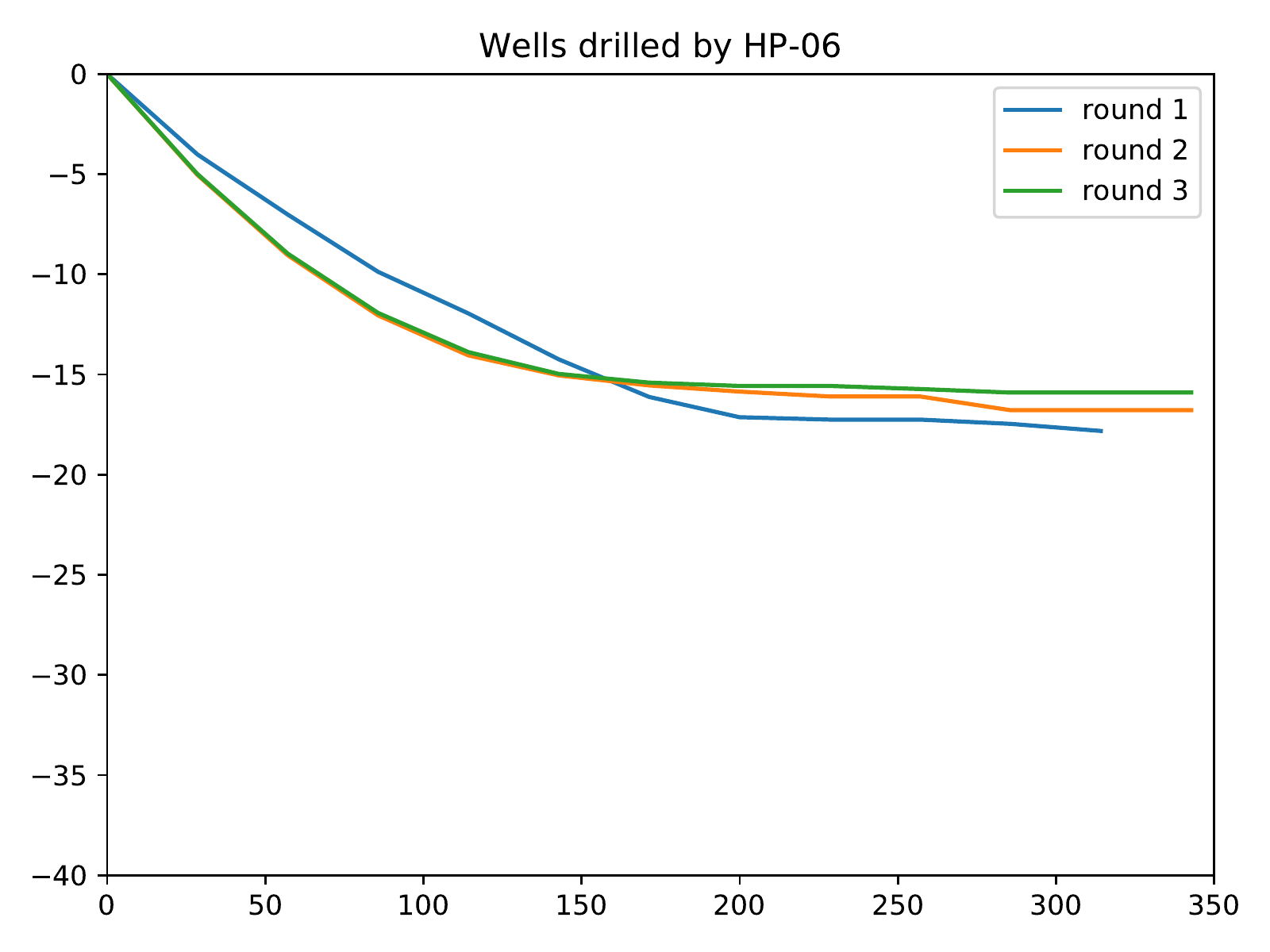}\, \quad\\
    c.
    \includegraphics[width=0.4\textwidth]{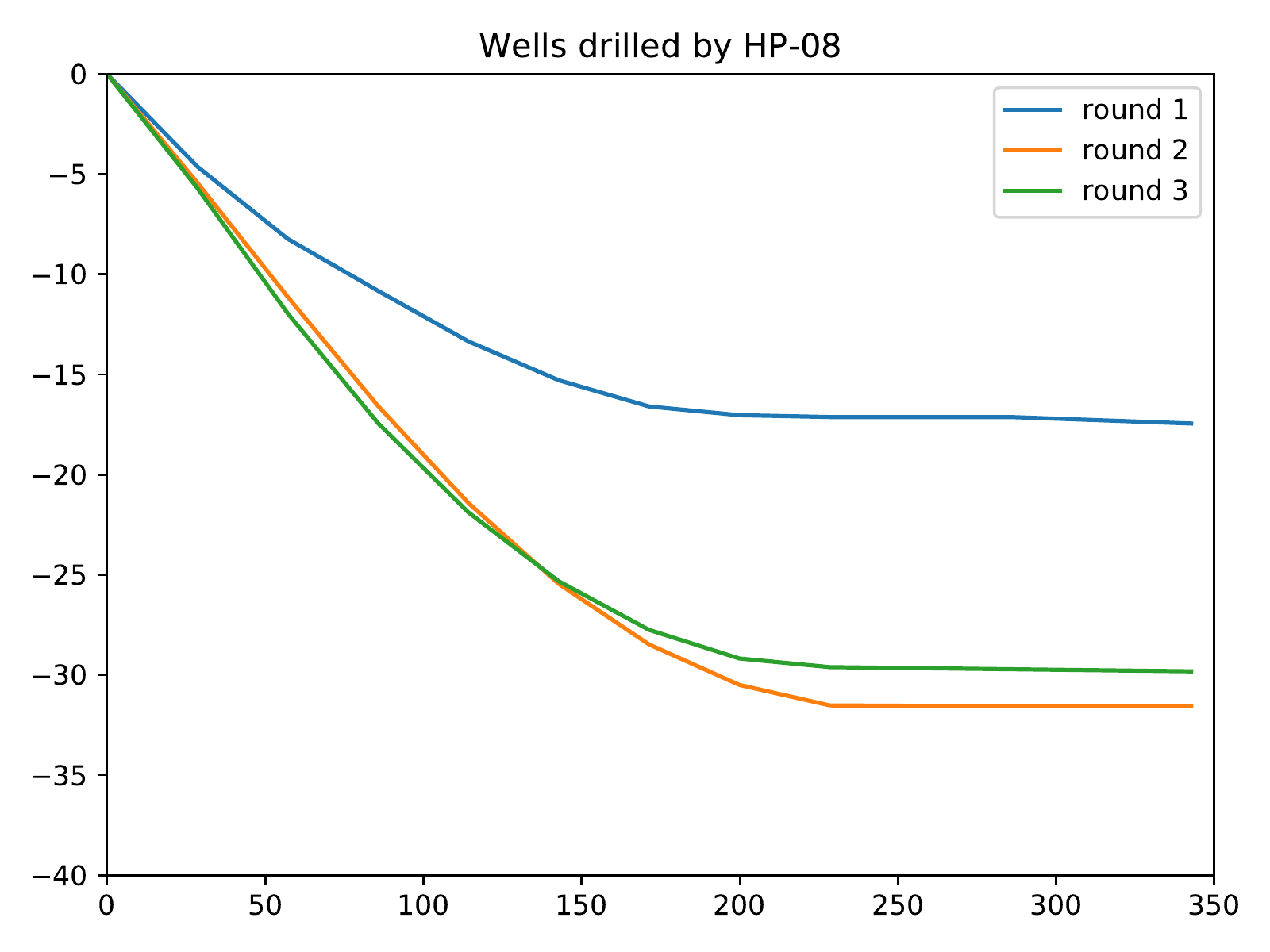}\,
    d.
    \includegraphics[width=0.4\textwidth]{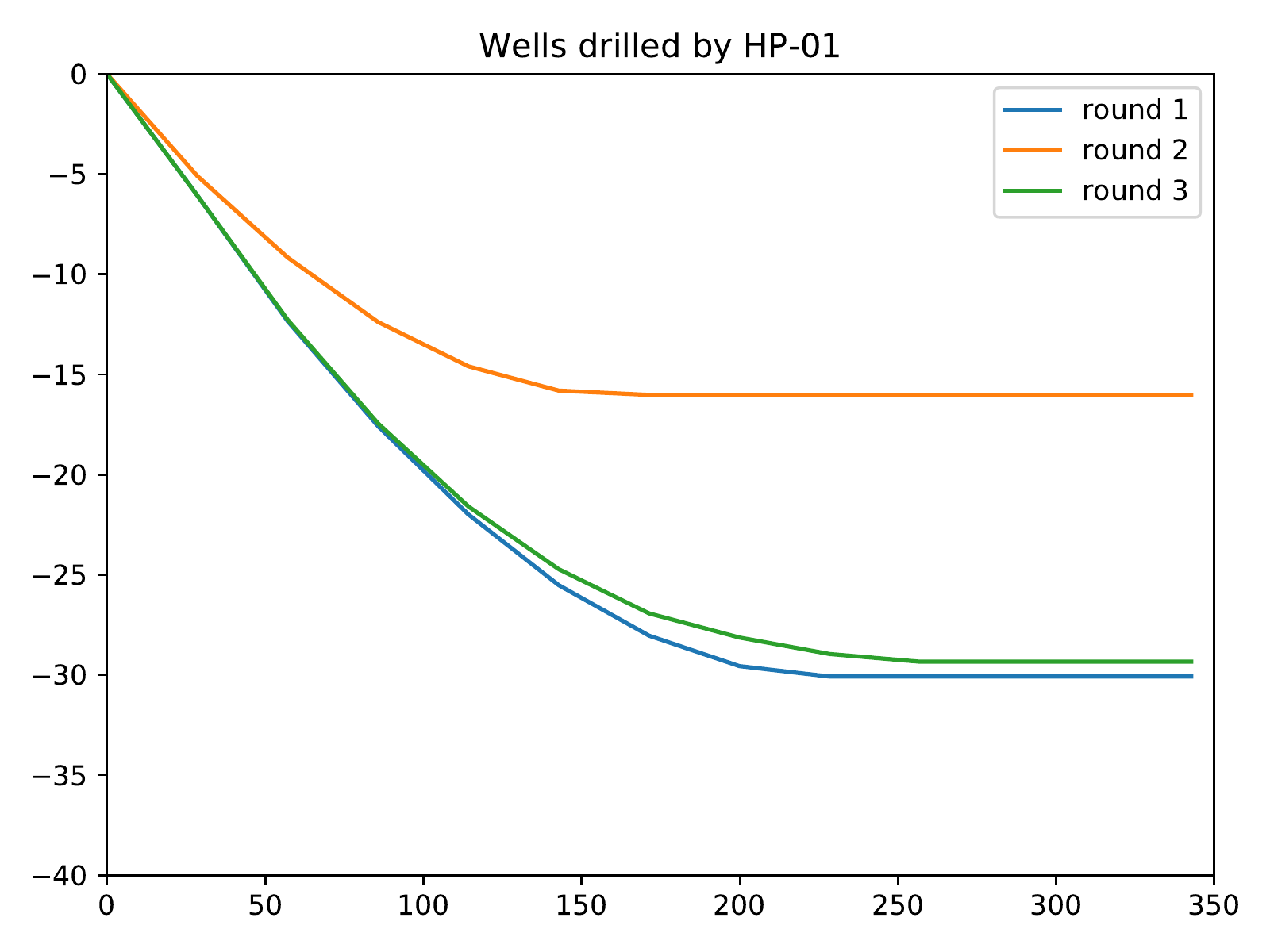}
    \caption{Examples of trajectories drilled by participants over the three rounds. 
    The rounds 2 and 3 have the same test set up, where the best solution was to drill into the top layer. 
    Round 1 has the other test, where the best solution was to drill into the bottom layer.
    Consistent (predictable) decision making results in identical or similar results for round 2 and 3.
    Each plot showing the best participant in its consistency group (see Figure \ref{fig:consistency}): a. absolutely consistent; b. consistent; c. relatively consistent; d. other. 
    }
    \label{fig:consistencyClassExamples}
\end{figure}

The automated DSS is built to ensure consistency. 
That is, the system is guaranteed to make the same decisions given the same input parameters.
This means repeatability  and hence predictability of automated decisions. 
The same perfect consistency would be impossible for human participants due to various factors, including the GUI limitations.
The experiment allowed us to test the extent to which the HPs were consistent in their decision making by comparing their decisions in the identical rounds 2 and 3.
Figure \ref{fig:consistencyClassExamples} shows the trajectories drilled by several participants with different levels of consistency:
\begin{itemize}
    \item \dss, which produces identical trajectories for identical set-ups.
    \item Consistent users, for whom the distance between trajectories for the same set-up was less than 0.5 meters on average.
    \item Relatively consistent users, for whom the consistency was worse than for consistent users, but distance between the trajectories in the same test was at least $2std$ lower than between different cases. Note that $std$ here is the standard deviation of average distance between the pairs of trajectories based on all combinations of tests 1-3.
    \item Other users for whom the consistency for the same set-ups was not observed.
\end{itemize}

All the results arranged by the level of consistency are shown in Figure \ref{fig:consistency}. 
For comparison the figure shows gray area of selecting the optimal layer purely by chance. 
\footnote{By guessing, one has 50\% chance to guess and aim for the optimal layer in each round. Given three rounds, a participant  has a 1/8th chance to aim for the optimal layer all three times.}
Thus, if all the participants did not use any relevant knowledge and tried to land and drill in a layer chosen randomly, about four of them should have selected the correct layer in all three rounds. 
From Figure \ref{fig:consistency} the number of consistent users is lower than probability of random guessing. 
The number including the relatively consistent users is still within possible error given relatively small number of total participants. 

Another important observation from 
Figure \ref{fig:consistency} is that the consistent users ended up with relatively low total score.
From this we can conclude that for HPs, the strategy that scored highest involved chance (early betting on which layer to land).
This is confirmed by the interview of the top-scoring participants \citep{alyaev2021systematic}. 
The interviews also reveal that the competition setting of the experiment made some of the participants alter their strategies towards more risk-taking compare to realistic geosteering.
Therefore the observed results do not necessarily reflect what experts would have done in a real operation.

\subsection{An interactive experiment as means of communication}
After the experiment, we asked the workshop participants to respond to an anonymous survey. 
The results in this section are based on responses from 21 participants.
The respondents rated the experiment as part of the workshop above 4 out of 5 on average.
The other questions were designed to evaluate the web-based-experiment platform itself as well as its applicability for training and communication of research.

In this subsection we present the feedback about usefulness of such an experiment for communication of the research. Connected to the experiment, we presented the ensemble-based geosteering workflow which was used behind the scenes in the web-based platform as well as the optimization behind the \dss.

\begin{figure}
    \centering
    \includegraphics[width=0.9\textwidth]{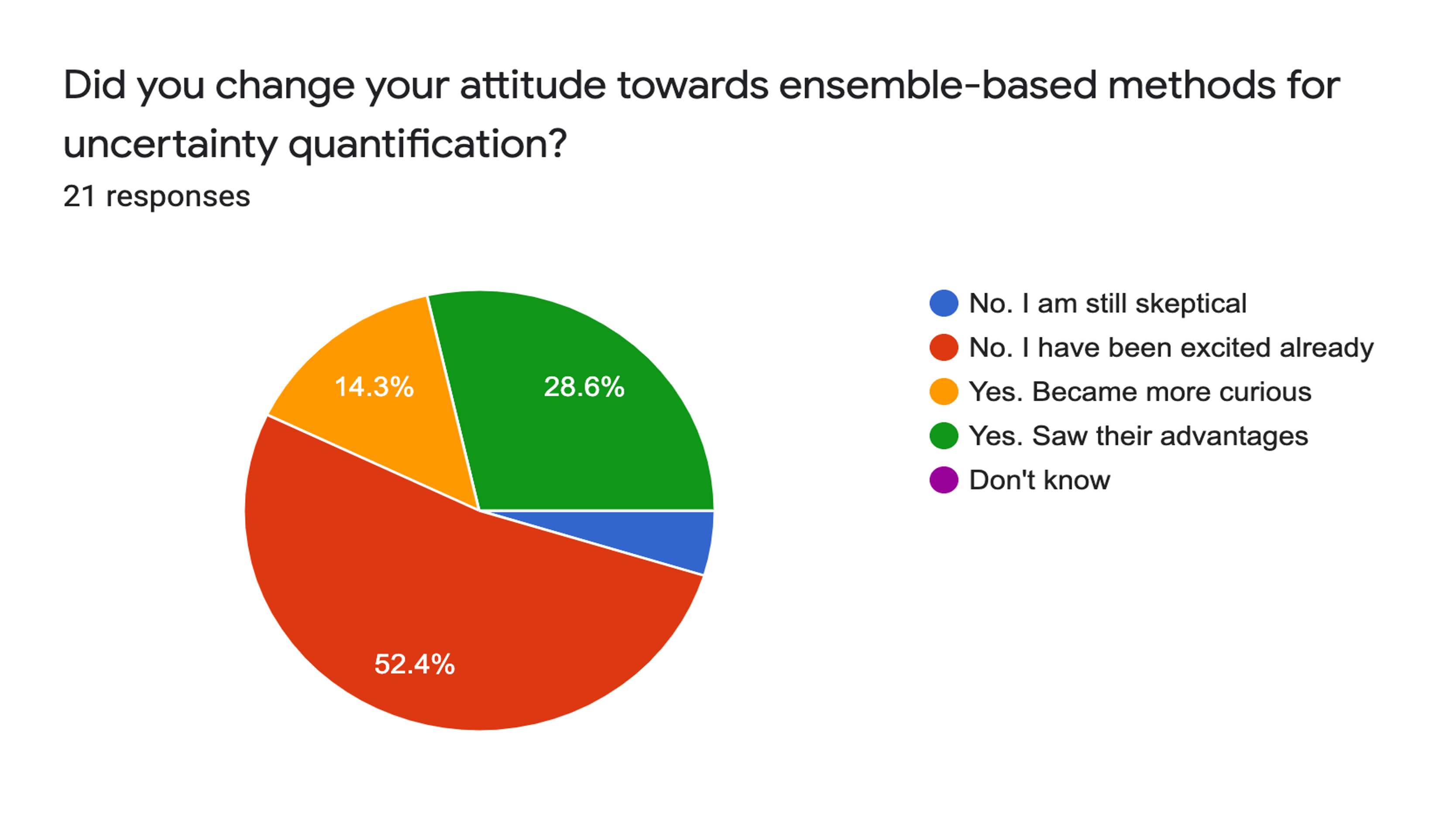}
    \includegraphics[width=0.9\textwidth]{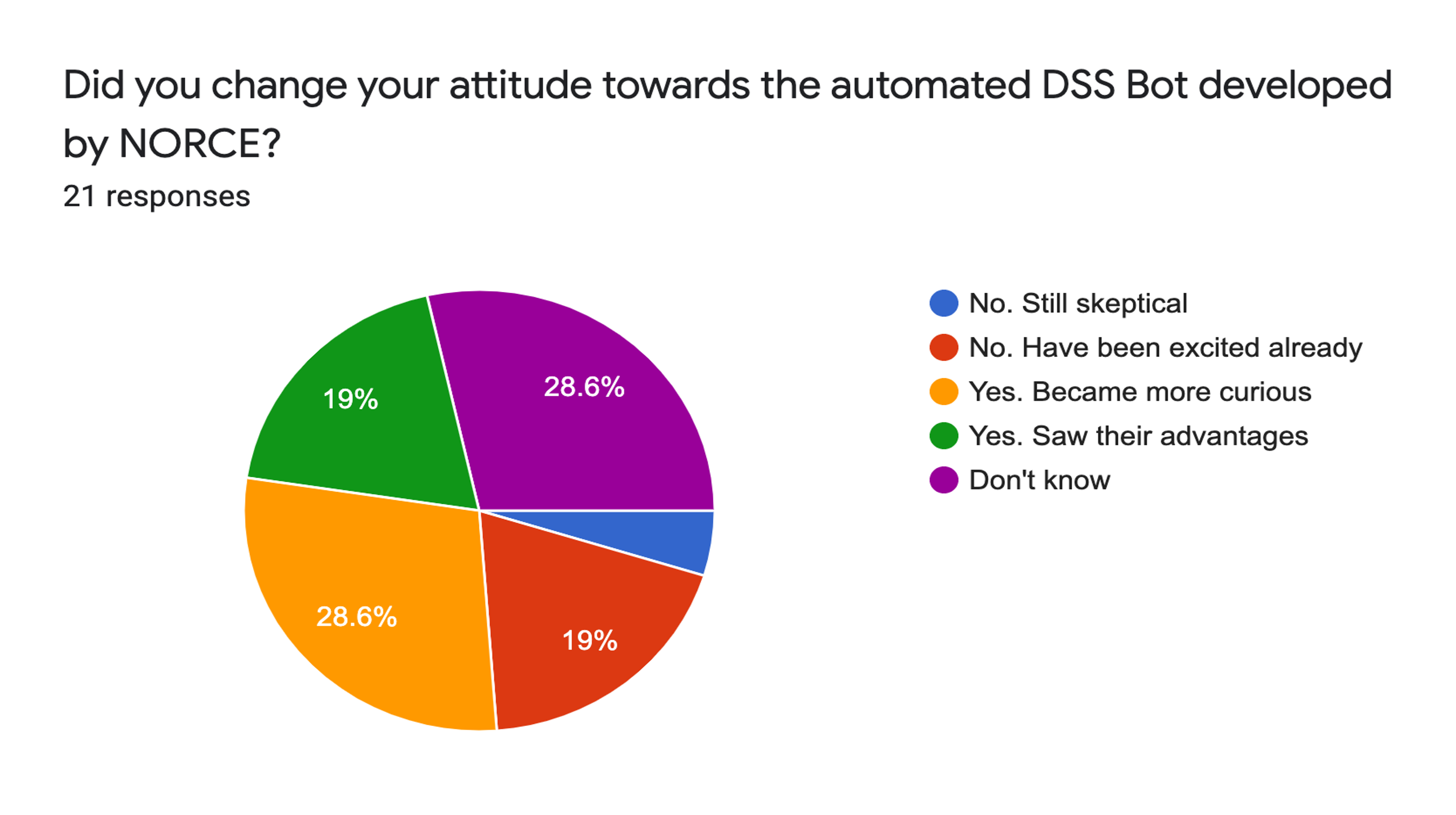}
    \includegraphics[width=0.9\textwidth]{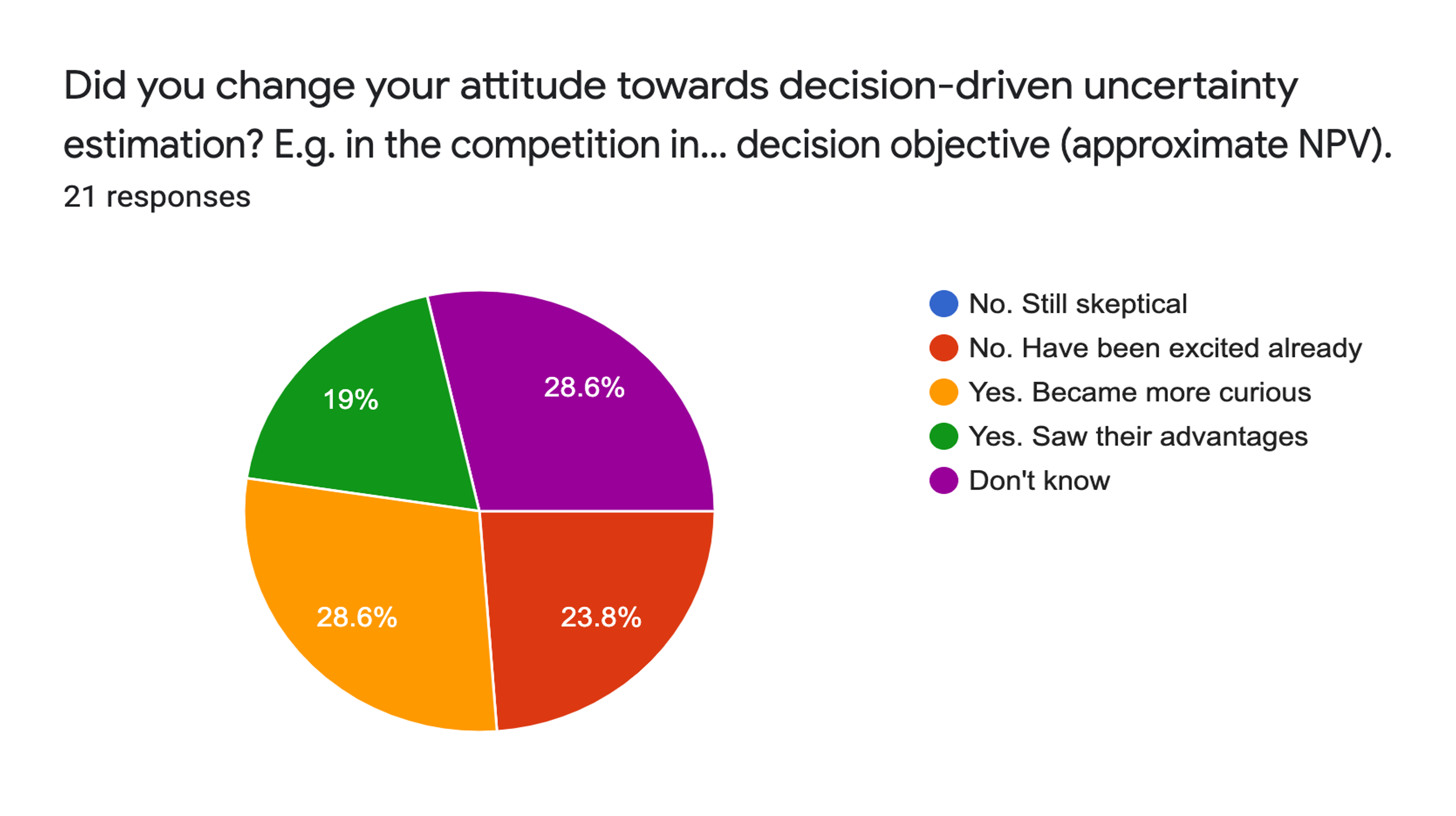}
    \caption{The value of the experiment in terms of communication of the research based on the survey.}
    \label{fig:research}
\end{figure}

In the survey we tried to see if the interactive experiment allowed us to make people more interested in the related research. Figure \ref{fig:research} shows whether and how the respondents changed there attitude towards the presented research concepts. 
The groups who were 'engaged' include those who became more curious and those who saw the advantages of the presented research.
We see that 47.6\% of respondents were engaged by the automated decision making and decision-driven uncertainty evaluation.
The lower 42.9\% engagement for the ensemble-based uncertainty quantification can be related to the much higher percentage (52.4\%) of the respondents who were already excited about these technologies. 
While we lack comparative data for standard workshop presentations, the communication-by-participation seems very powerful based on the results of the survey. 

\subsection{User feedback to web-based platform}
It is known that the uncertainty visualization influences decision making \citep{Viard2011}.
Therefore, in the second part of the questionnaire we asked the participants to evaluate the usefulness of different elements of the platform to allow further improvements of the platform.  
This included the elements of the GUI, but also the ability to use the platform from any browser-equipped device. 

\begin{figure}
    \centering
    \includegraphics[width=0.9\textwidth]{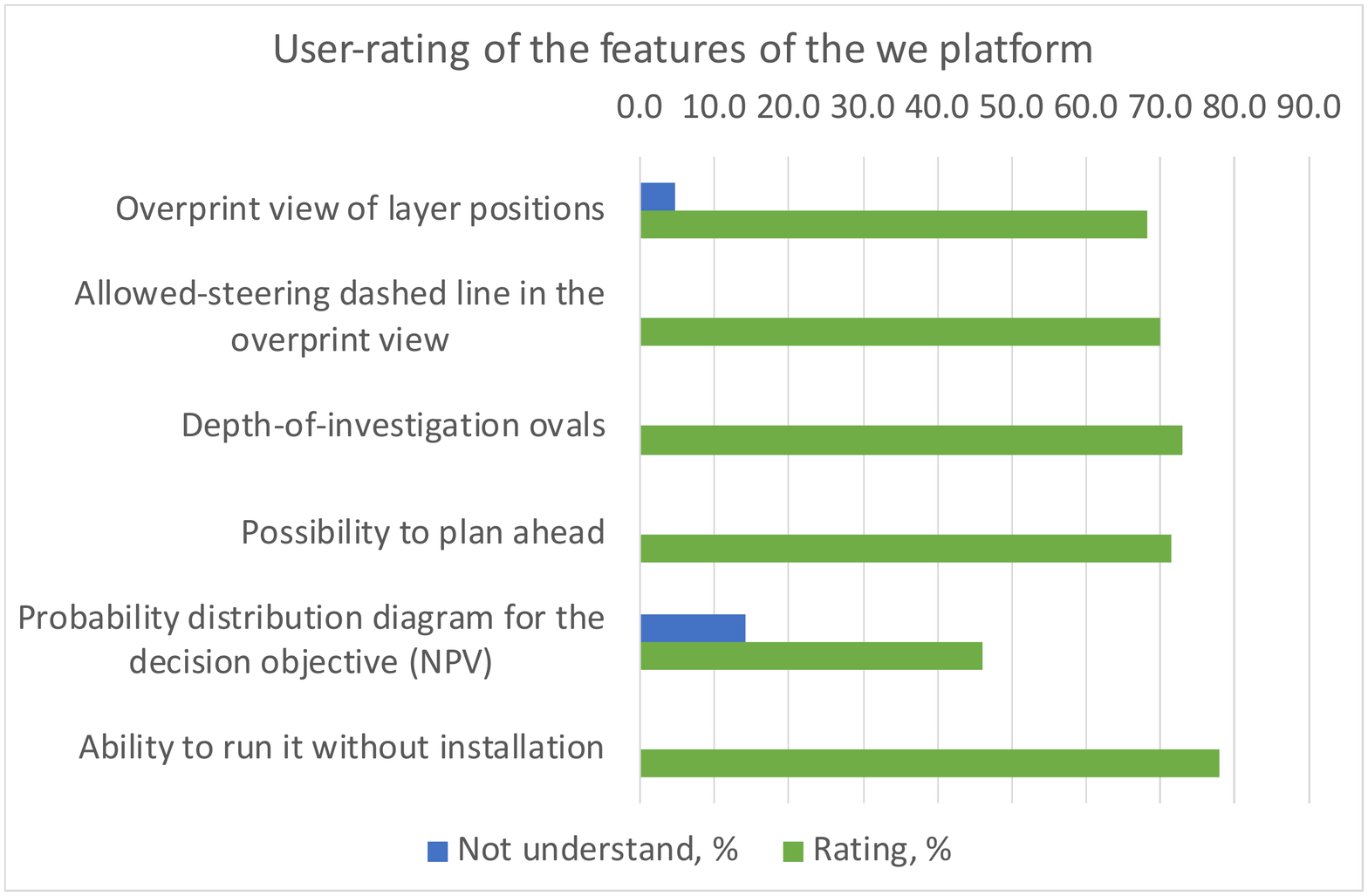}
    \caption{Relative importance of the features of the web-based platform according to the conducted survey. 100\% corresponds to all respondents considering a feature "Very useful". The diagram also shows the percentage of respondents who did not understand a feature.}
    \label{fig:inerfaceFeatures}
\end{figure}

The respondents evaluated every feature of the platform on the following scale:
\begin{itemize}
    \item I did not understand it
    \item[0] Not useful
    \item[1] Somewhat useful
    \item[2] Quite useful
    \item[3] Very useful
\end{itemize}
Figure \ref{fig:inerfaceFeatures} shows relative importance of each feature as the sum of the scores by the rules above in percent. I.e. 100\% correspond to all participants answering "Very useful". It also shows the fraction of the participants who did not understand a feature.

The ability to run the GUI without installation has the highest relative importance. 
This supports our design decision to develop the web-based GUI for the experimental platform as a paradigm for user-based experiments.

Looking at the rest of the distribution in Figure \ref{fig:inerfaceFeatures}, we observe that most features of the GUI have rating around 70\%. 
This indicates that that overall communication of the GUI during the experiment was successful despite its relatively short format.
The lowest scores correspond to probability distribution and the overprint display. 
These two features of the GUI were designed to communicate the uncertainty, which is not straight-forward. 
Further experimental studies are required to improve the uncertainty communication.

\begin{figure}
    \centering
    \includegraphics[width=0.9\textwidth]{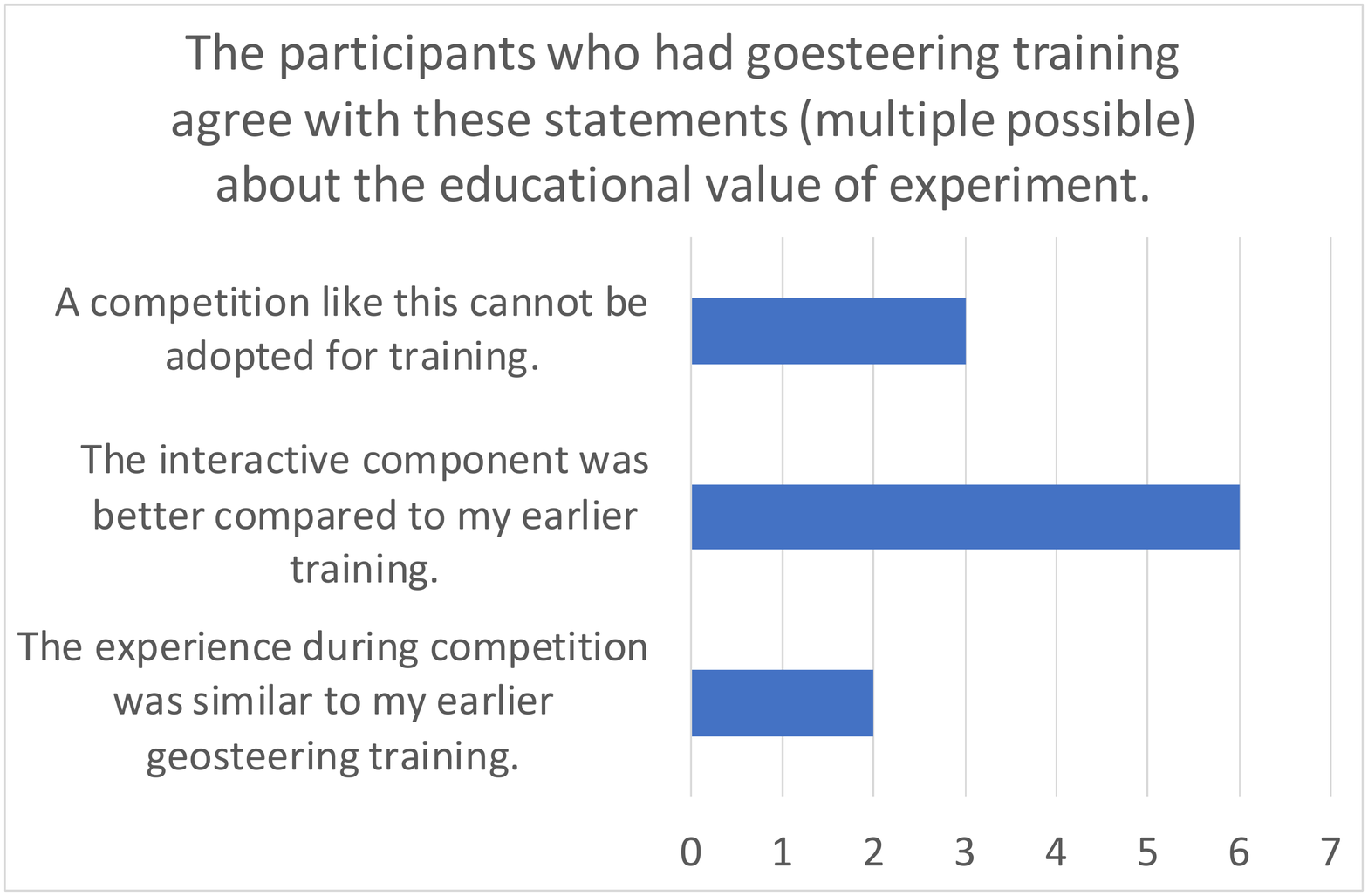}
    \caption{Value of the web-based platform for geosteering training.}
    \label{fig:forTraining}
\end{figure}

Considering that the web-based platform was overall useful for the experiment, we also asked how such an experiment would stand against the current training practices in geosteering. 
Figure \ref{fig:forTraining} shows that most of the respondents think that the interactive component is similar (2) or better (6)  compare to standard training practices.  
At the same time, a fraction (3) think that the experiment cannot be adopted for training in geosteering. 

\section{Conclusions}
\label{sec:conclusions}

In this paper we have presented a web-based platform which provides users with the opportunity to perform assisted decision-making under uncertainty in a benchmark environment.
While there has been previous benchmark environments made for development and training AI agents \citep{brockman2016openaigym},
to our knowledge the presented platform is novel in the geoscience/geosteering context.
Moreover, it provides the possibility to compare the results of decision making by human experts with automated algorithms providing similar information to both.

In this paper we present the first experiment, which put 29 geoscientists against the DSS algorithm from \cite{alyaev2019dss}. 
The results show that \dss{} outperformed all but one qualified participants considering relative wells' value
(doing better than 94\%).
What's more, no participant beat \dss{} more than once over the three rounds, giving it the best comparative rating among the participants.

Based on the data from experiment we were able to evaluate experts' strategies and measure the consistency of their decisions. 
Only two experts were consistent within the tolerance specified, but achieved much lower total score relative to \dss{} which is consistent by design. 
However, according to an interview \citep{alyaev2021systematic} the strategies of participants could have been affected by the artificial setting of the experiment and thus cannot yield conclusions about operational decision making.

The feedback collected from the participants indicates that the experiment provided both entertaining and educational value. 
The respondents particularly appreciated the ability to run the experiment without any installation and the interactivity of the setup. 
As a result 76\% of respondents became more curious and/or saw the advantages of the technologies related to the experiment. 
These include ensemble-based representation and updating of uncertainty, decision-driven uncertainty estimation, and technologies behind \dss \citep{alyaev2019dss}.
Thus, we can conclude that the platform is valuable as medium for research communication.

The results of the experiment  gives better understanding of the specific challenges in the human-system interaction.
For the presented platform future work will include improvement of visualization and/or communication of uncertainty of the well value in the score distribution diagram.
This knowledge is essential 
for successful further development and  adoption of decision support systems in geosciences. 
The proposed platform and its derivatives 
can serve 
as a benchmark which will aid training of people and development of algorithms.




\section{Acknowledgments}
We thank Robert Ewald for help with deployment of the platform for the experiment.

Funding:  This work was supported by the research project 'Geosteering for IOR' (NFR-Petromaks2 project no. 268122) which is funded by the Research Council of Norway, Aker BP, Equinor, V{\aa}r Energi and Baker Hughes Norway.

\section{Computer Code Availability}
\label{sec:repo}
The web-based API and GUI described in this paper is developed by the authors of the paper in 2019 and, since October 2020, is available as a stand-alone repository  
"API and GUI for GEOSTEERING benchmark the NORCE way" 
at
\url{https://github.com/NORCE-Energy/geosteering-game-gui} under the MIT license.
The code contains a reference implementation of a decision agent in Python 3 (API), and the Python server that runs a copy of the GUI of the web-based in JavaScript/HTML/CSS locally. 
The code has been tested with Python 3.7 and Google Chrome	86.0.4240.111 
on 	macOS v.10.15.7 and Windows 10 v.1709.

The server back-end uses the library described in \cite{alyaev2019dss}, which is currently proprietary to NORCE, but NORCE is committed to maintain the availability of the back-end to facilitate further development of the GUI and creation of new decision agents within the benchmark. 
The link to the available back-end servers will be update in the repository.

The repository also contains the files with the results of the anonymized results of first experiment described in Section \ref{sec:experiment} and a script for their playback.

For any questions related to the repository, please, contact the corresponding author: Sergey.Alyaev@norceresearch.no, +47 518 75 610.

\bibliography{library,manual-bib}





\end{document}